\documentclass[onecolumn]{emulateapj}  
\usepackage{epstopdf}
\usepackage{ulem}
\usepackage{amssymb}
\usepackage{natbib}
\usepackage{times}
\usepackage{graphicx}
\usepackage[usenames,dvipsnames]{pstricks}
 \usepackage{psfrag}
                                                                                        
\usepackage[colorlinks=true,linkcolor=blue,citecolor=blue]{hyperref}

\def\aj{AJ}
\def\araa{ARA\&A}
\def\apj{ApJ}
\def\apjl{ApJ}
\def\apjs{ApJS}
\def\apss{Ap\&SS}
\def\aap{A\&A}
\def\aaps{A\&AS}
\def\jcap{J. Cosmology Astropart. Phys.}
\def\mnras{MNRAS}
\def\prd{Phys.~Rev.~D}
\def\pasj{PASJ}
\def\ssr{Space~Sci.~Rev.}
\def\nat{Nature}
\def\physrep{Phys.~Rep.}

\newcommand{\be}{\begin{equation}}
\newcommand{\ee}{\end{equation}}
\newcommand{\bary}{\begin{eqnarray}}
\newcommand{\eary}{\end{eqnarray}}
\newcommand{\en}{E_\nu}
\newcommand{\enc}{E_{\nu,c}}

\shorttitle{Neutrinos, $\gamma$-rays and cosmic rays from LL AGN }
\shortauthors{Fraija et al.}
\begin{document}
\title{Neutrino, $\gamma$-ray and cosmic ray fluxes from the core of the\\ closest radio galaxies}
\author{N. Fraija\altaffilmark{1} and  A. Marinelli\altaffilmark{2}}
\affil{$^1$Instituto de Astronom\'ia, Universidad Nacional Aut\'{o}noma de M\'{e}xico, Apdo. Postal 70-264, Cd. Universitaria, DF 04510, M\'{e}xico}
\affil{$^2$ Dipartimento di Fisica, Universita di Pisa and I.N.F.N., Largo Bruno Pontecorvo, 3, 56127 Pisa, Italia}
\email{nifraija@astro.unam.mx, antonio.marinelli@pi.infn.it}
\date{\today} 

\begin{abstract}
The closest radio galaxies;  Centaurus A, M87 and NGC 1275,  have been detected from radio wavelengths to TeV $\gamma$-rays, and also studied as high-energy neutrino and ultra-high-energy cosmic ray potential emitters.  Their spectral energy distributions show a double-peak feature, which is explained by synchrotron self-Compton model. However,  TeV $\gamma$-ray measured spectra could suggest that very-high-energy $\gamma$-rays might have a hadronic origin. We introduce a lepto-hadronic model to describe the broadband spectral energy distribution;  from radio to sub GeV photons as  synchrotron self-Compton emission and  TeV $\gamma$-ray photons as neutral pion decay resulting from p$\gamma$ interactions occurring close to the core. These photo-hadronic interactions take place when Fermi-accelerated protons interact with the seed photons around synchrotron self-Compton peaks.  Obtaining a good description of the TeV $\gamma$-ray fluxes,  firstly, we compute neutrino fluxes and events expected in IceCube detector and secondly,  we estimate ultra-high-energy cosmic ray fluxes and event rate expected in Telescope Array, Pierre Auger and HiRes observatories.  Within this scenario we show that the expected high-energy neutrinos cannot explain the astrophysical flux observed by IceCube, and the connection with ultra-high-energy cosmic rays observed by Auger experiment around Centaurus A, might be possible only considering a heavy nuclei composition in the observed events.
\end{abstract}


\section{Introduction}

%
Radio galaxies (RGs) are radio loud active galaxy nuclei (AGN) exhibiting clear structure of a compact central source,  large-scale jets and lobes.  These  sources are also of interest due to the close proximity to the earth affording us an excellent opportunity for studying the physics of relativistic outflows.  They  have been widely studied from radio wavelengths to  MeV-GeV $\gamma$-rays and recently at very-high-energy (VHE) by Imagine Atmospheric Cherenkov Telescope (IACT).  RGs are generally described by the standard non-thermal  synchrotron self-Compton (SSC) model \citep{2010ApJ...719.1433A,2012ApJ...753...40F,1998ApJ...509..608T}.  In SSC framework,  low-energy emission; radio through optical, originates from synchrotron radiation while HE photons; X-rays through $\gamma$-rays, come from inverse Compton scattering emission.   However, this model with only one population of electrons,  predicts a spectral energy distribution (SED) that can be hardly extended up to higher energies than a few GeVs \citep{2005ApJ...634L..33G,2008A&A...478..111L}. In addition, some authors have suggested that the emission in the GeV - TeV energy range may have origins in different physical processes \citep{2011MNRAS.413.2785B,2005ApJ...634L..33G}.\\
%
%
The Telescope Array  (TA) experiment, located in Millard Country (Utah), was designed to study ultra-high-energy cosmic rays (UHECRs) with energies above 57 EeV \citep{2012NIMPA.689...87A}.  TA observatory, with a field of view covering the sky region above -10$^\circ$ of declination, detected a cluster of 72 UHECRs events centered at R. A.=146$^\circ$.7, Dec.=43$^\circ.2$.  It had a Li-Ma statistical significance of 5.1$\sigma$ within 5 years of data taking \citep{2014arXiv1404.5890T}.   The Pierre Auger observatory (PAO), located in the Mendoza Province of Argentina, was designed to determine the arrival directions and energies of UHECRs using four fluorescence telescope arrays and the High Resolution Fly's eye (HiRes) experiment, located in the west-central Utah desert, consisted of two detectors that observed cosmic ray showers via the fluorescence light.  PAO studying the composition of the high-energy showers  found that the distribution of their properties was located in somewhere between pure proton (p) and pure iron (Fe) at 57 EeV\citep{2008ICRC....4..335Y, 2008APh....29..188P, 2007AN....328..614U}, although the latest results favored a heavy nuclear composition \citep{2010PhRvL.104i1101A}.  By contrast,  HiRes data were consistent with a dominant proton composition at this energy range \citep{2007AN....328..614U, 2008ICRC....4..385E}. PAO detected 27 UHECRs,  with two of them associated with Centaurus A (Cen A), being reconstructed inside a circle centered at the position of this FR I with a aperture of 3.1$^{\circ}$.\\
%
%
The IceCube detector located at the South Pole was designed to record the interactions of neutrinos. Encompassing a cubic kilometer of ice and almost four years of data taking (from 2010 to 2014), the IceCube telescope reported with the High-Energy Starting Events (HESE)\footnote{http://icecube.wisc.edu/science/data/HE-nu-2010-2014} catalog a sample of 54 extraterrestrial neutrino events in the TeV - PeV energy range.   Arrival directions of these events are compatible with an isotropic distribution. The neutrino flux is compatible with a high component due to extragalactic origin \citep{2014PhRvD..90b3010A, 2015ApJ...815L..25G}. For instance, gamma-ray bursts \citep{2013PhRvL.111l1102M, 2013PhRvD..88j3003R, 2014MNRAS.437.2187F, 2013ApJ...766...73L, 2015JCAP...09..036T, 2014MNRAS.445..570P, 2016JHEAp...9...25F} and AGN (\cite{2014JHEAp...3...29D,2013PhRvD..88d7301S, 2015MNRAS.448.2412P, 2015MNRAS.452.1877P, 2014arXiv1410.8549M, 2015APh....71....1F, 2015APh....70...54F}), etc.\\
%
%
Hadronic processes producing VHE neutrinos and photons through the acceleration of cosmic rays in AGNs have been explored by many authors \citep{2001PhRvL..87v1102A, 2008PhR...458..173B, 2014arXiv1410.8124K, 2001APh....15..121M, 2008PhRvD..78b3007C,  2009NJPh...11f5016D, 2015MNRAS.451.1502T, 2015ApJ...806..159K, 2007ApJ...670L..81H, 2003ApJ...593..169H}.  In particular,  GeV-TeV $\gamma$-ray fluxes interpreted as pion decay products from photo-hadronic interactions occurring close to the core of RGs have been also discussed for different sources \citep{2012PhRvD..85d3012S, 2012ApJ...753...40F, 2015arXiv150407592K, 2016MNRAS.455..838K, 2014MNRAS.441.1209F, 2014A&A...562A..12P,2015IAUS..313..175F}.\\
%
%
In this work we introduce a leptonic and hadronic model to describe the broadband SED of the closest RGs.  For the leptonic model,  we present the SSC model to explain the SED up to dozens of GeV and for the hadronic model, we propose that the proton-photon (p$\gamma$) interactions occurring close to the core could describe the $\gamma$-ray spectrum at the GeV - TeV energy range.  Correlating the TeV $\gamma$-ray, UHECR and neutrino spectra through p$\gamma$ interactions, we estimate their fluxes and number of events expected in PAO, TA and HiRes experiments and IceCube telescope, respectively. For this correlation, we have assumed that the proton spectrum is extended through a simple  power law up to UHEs.
\section{The closest RGs: Cen A, M87 and NGC1275}
In this section, we are going to present a brief description of the three RGs studied in this work and the set of data used in our analysis.
\subsection{Cen A}
Cen A, at a distance of d$_z\simeq3.8$ Mpc (z=0.00183), has been one of the best studied radio galaxies.  It is characterized by having an off-axis jet estimated in $\sim 15^\circ - 80^\circ$ \citep[see, e.g.] [and reference therein]{2006PASJ...58..211H} and two giant radio lobes.   Cen A has been imaged in radio wavelengths, optical bands \citep{1975ApJ...199L.139W,1976ApJ...206L..45M,1970ApJ...161L...1B,1981ApJ...244..429B},  X-rays, sub-GeV$\gamma$-rays and VHE $\gamma$-rays \citep{2003ApJ...593..169H,1999APh....11..221S,2009ApJ...695L..40A, 2010ApJ...719.1433A, 1998A&A...330...97S, 2010ApJ...719.1433A,2013ApJ...770L...6S}.  This radio galaxy was observed during a period of 10 months by the Large Area Telescope (LAT) on board the Fermi. The sub-GeV $\gamma$-ray flux collected  was described with  a power law (F$\propto E^{-\alpha}$ with $\alpha= 2.67\pm 0.10_{\rm stat} \pm 0.08_{\rm sys}$; \cite{2010ApJ...719.1433A}).  In addition,  Cen A was also detected for more than 120 hr by the High Energy Stereoscopic System (HESS) \citep{2005A&A...441..465A, 2009ApJ...695L..40A}.  The observed spectrum ($>$ 250 GeV) was fitted using a simple power law with a spectral index of 2.7$\pm\, 0.5_{\rm stat} \pm 0.2_{\rm sys}$ and an apparent luminosity of $\simeq 2.6 \times 10^{39}$ erg s$^{-1}$.  No significant flux variability was detected.
\subsection{M87}
M87, at a distance of d$_z\simeq16$ Mpc (z=0.0043), is located in the Virgo cluster of galaxies and hosts a central black hole (BH) of $(3.2\pm 0.9)\times 10^9$ solar masses \citep{1996ApJ...467..597B}.  The jet, inclined at an angle of $\sim$ 15$^\circ$ - 20$^\circ$ relative to the observe's line of sight, has been imaged in its base down to $\sim$0.01 pc resolution \citep{1999Natur.401..891J, 2007ApJ...660..200L}.     As one of the nearest radio galaxies to us, M87 is amongst the best-studied of its source class. It has been detected from radio wavelengths to VHE $\gamma$ rays.   Above 100 MeV,  M87 has been  detected by LAT-Fermi \citep{2009ApJ...707...55A}, HESS, MAGIC and the Very Energetic Radiation Imaging Telescope Array System (VERITAS) telescopes \citep{2012ApJ...746..151A, 2009ApJ...706L.275A, 2012ApJ...746..141A}.  In 2004, the HESS telescope detected the active galaxy M87 in its historical minimum. The TeV flux measurement was well fitted with a simple power law.  The differential energy spectrum was  $(2.43\pm0.75)\times10^{-13} \,\, {\rm TeV^{-1}\,cm^{-2}\,s^{-1}}$ with a photon index of $2.62\pm0.35$ \citep{2006Sci...314.1424A}.    No indications for short-term variability were found.
\subsection{NGC1275}
NGC 1275, also known as Perseus A and 3C 84, is an elliptical/radio galaxy located at the center of the Perseus cluster at d$_z\simeq76$ Mpc ($z=0.0179$).   This source has a strong, compact nucleus which has been studied in detail with Very Long Baseline Interferometry (VLBI)\citep{1994ApJ...430L..41V, 2006MNRAS.368.1500T, 2000ApJ...530..233W, 2006PASJ...58..261A} and Array (VLBA) \citep{2014ApJ...785...53N}.  These observations reveal a compact core and a bowshock-like souther jet component moving steadily outwards at 0.3 mas/year \citep{2004ApJ...609..539K,2009AJ....137.3718L}.  The norther counter jet is also detected, though it is much less prominent due to Doppler dimming, as well as to free-free absorption due an intervening disk.    Walker, Romney and Berson (1994) derive from these observations that the jet has an intrinsic velocity of $0.3c - 0.5c$ oriented at an angle $\approx$ 10$^\circ$ - 35$^\circ$ to the line of sight.  Polarization has recently been detected in the southern jet \citep{2006MNRAS.368.1500T}, suggesting increasingly strong interactions of the jet with the surrounding environment.\\   
Due to its brightness and proximity, this source has been detected from radio to TeV $\gamma$-ray bands \citep{2009ApJ...699...31A}.   This radio galaxy was observed at energies above 100 MeV by Fermi-LAT from 2008 August 4 to 2008 December 5.  The average flux and photon index measured, which remained relatively constant during the observing period were $(2.10\pm0.23)\times 10^{-7}\, {\rm  ph\, cm^{-2}\, s^{-1}}$ and $2.17\pm 0.05$, respectively.    This source has been detected by MAGIC telescope with a statistical significance of $6.6\,\sigma$ above 100 GeV in 46 hr of stereo observations carried out between August 2010 and February 2011.  The measured differential energy spectrum between 70 GeV and 500 GeV was described by a power law with a steep spectral index of $-4.1\pm0.7_{stat}\pm0.3_{\rm sys}$, and an average flux of $(1.3\pm 0.2_{stat}\pm0.3_{\rm sys})\times 10^{-11}\,{\rm cm^{-2}\,s^{-1}}$ \citep{2012A&A...539L...2A}. The light curve  ($>$ 100 GeV) did not show hints of variability on a month time scale. 
\section{Theoretical Model}
We propose that the broadband SED of RGs can be described as the superposition of SSC emission and $\pi^0$ decay product resulting from p$\gamma$ interactions.
\subsection{Leptonic Model}
Fermi-accelerated electrons are injected in an emitting region with radius $r_d$ which moves at ultra-relativistic velocities with Doppler factor $\delta_D$.  Relativistic electrons are confined in the emitting region  by a magnetic field, therefore it is expected non-thermal photons by synchrotron and Compton scattering radiation. 
\subsubsection{Synchrotron radiation}
The electron population   is described by a broken power-law given by \citep{1994hea2.book.....L}
\begin{equation}
\label{espele}
N_e(\gamma_e)   = N_{0,e}
\cases {
\gamma_e^{-\alpha_e} 						& 	$\gamma_{\rm e,min}<\gamma_e < \gamma_{\rm e,c}$,\cr
\gamma_{e,c}    \gamma_e^{-(\alpha_e+1)}          & 	$\gamma_{\rm e,c} \leq  \gamma_e<\gamma_{\rm e,max}$,\cr
}
\end{equation}
\noindent where  $N_{0,e}$ is the proportionality electron constant, $\alpha_e$ is the spectral power index of the electron population and $\gamma_{e,i}$ are  the electron Lorentz  factors. The index $i$ is min, c or max for minimum, cooling and maximum, respectively.   By considering that a fraction of total  energy density is given to accelerate electrons  $U_e=m_e \int\gamma_e N_e(\gamma_e)d\gamma_e$, then the minimum electron Lorentz factor and the electron luminosity can be written as
\bary\label{gamma_m}
\gamma_{\rm e,min}&=&\frac{(\alpha_e-2)}{m_e(\alpha_e-1)}\,\frac{U_e}{N_e}\,,
\eary
and
\be\label{Le}
L_e=4\pi\, \delta^2_D\,r^2_d\,U_e\,,
\ee
respectively,  where $m_e$ is the electron mass.   The electron distribution in the emitting region permeated by a magnetic field $B=\sqrt{8\pi\,U_B}$ cools down following the cooling synchrotron time scale $t'_c=\frac{3m_e}{4\sigma_T}\,\,U^{-1}_B\,\gamma^{-1}_e$, with  $\sigma_T=6.65 \times 10^{-25}\, {\rm cm^2}$ the Compton cross section.  Comparing the synchrotron time scale with the dynamic scale $t'_d\simeq r_d/\delta_D$, we get that the cooling Lorentz factor is 
\be\label{gamma_b}
\gamma_{\rm e,c}= \frac{3\,m_e}{4\,\sigma_T}\,(1+Y)^{-1}\, \delta_D\, U_B^{-1}\,r_d^{-1}\,,
\ee
where the Compton parameter is
{\small
\be\label{Y1}
Y\equiv\frac{L_{\rm \gamma, IC}}{L_{\rm syn}}=\frac{U_{\rm rad}}{U_B}=\frac{U_{\rm syn}}{U_B}=\frac{\eta\, U_e}{(1+Y)\,U_B}\,.
\ee
}
Here, $L_{\rm \gamma, IC}$ is the observed luminosity around the second SSC peak,  $U_{\rm syn}$ and $U_B$ are the energy density of synchrotron radiation and magnetic field, respectively. Solving equation (\ref{Y1}), the two interesting limits are
{\small
\begin{eqnarray}
\label{Y2}
Y=\cases{ 
\frac{\eta\,U_e}{U_B}\,,     & if  $\frac{\eta\,U_e}{U_B}\ll 1$   ,\cr
\left(\frac{\eta\,U_e}{U_B}\right)^{1/2}\,,  & if  $\frac{\eta\,U_e}{U_B}\gg 1$,\cr
}
\end{eqnarray}
}
with $\eta=(\gamma_{\rm e,c}/\gamma_{\rm e,min})^{2-\alpha_e}$ given for slow cooling and $\eta=1$ for fast cooling  \citep{2001ApJ...548..787S}.  By considering that acceleration time scale $t'_{acc}\simeq\sqrt\frac\pi2m_e/q_e\, U^{-1/2}_B\,\gamma_e$ and cooling time scale are similar, it is possible to write the maximum electron Lorentz factor as
\bary\label{gamma_max}
\gamma_{\rm e,max}&=&\biggl(\frac{9\, q_e^2}{8\pi\, \sigma_T^2}\biggr)^{1/4}\,U_B^{-1/4}\,,
\eary 
\noindent where $q_e$ is the electric charge.  Taking into account the synchrotron emission ${\small \epsilon_\gamma(\gamma_{e,i})=\sqrt{\frac{8\pi q_e^2}{m_e^2}}\,\delta_D\, U_B^{1/2}\, \gamma^2_{e,i}}$ and  eqs. (\ref{gamma_m}), (\ref{gamma_b}) and (\ref{gamma_max}),  the synchrotron break energies are
\begin{eqnarray}\label{synrad}
\epsilon^{\rm syn}_{\rm \gamma,m} &=& \frac{\sqrt{8\pi}\,q_e}{m_e}\,\delta_D\,U_B^{1/2}\,\gamma^2_{\rm e,min}\cr
\epsilon^{\rm syn}_{\rm \gamma,c} &=&\frac{ 9\sqrt{2\pi}\,q_e\,m_e}{8\,\sigma_T^2}\, (1+Y)^{-2}\,\delta_D^3\, U_B^{-3/2}\, r_d^{-2}\cr
\epsilon^{\rm syn}_{\rm \gamma, max} &=&\frac{3\,q_e^2}{m_e\,\sigma_T}\, \delta_D\,.
\end{eqnarray}
The synchrotron spectrum is computed  through the electron distribution  (eq. \ref{espele}) rather than synchrotron radiation of a single electron. Therefore, the photon energy radiated in the range $\epsilon_\gamma$ to $\epsilon_\gamma + d\epsilon_\gamma$ is given by electrons between   $E_e$ and $E_e + dE_e$; then we can estimate the photon spectrum through emissivity $\epsilon_\gamma N_\gamma(\epsilon_\gamma) d\epsilon_\gamma=(- dE_e/dt)\,N_e(E_e)dE_e$.  Following  \cite{1994hea2.book.....L} and \cite{1986rpa..book.....R}, one can show that if electron distribution has  spectral indexes  $\alpha_e$ and $(\alpha_e-1)$, then the photon distribution has  spectral indexes $(\alpha_e-1)/2$ and $\alpha_e/2$, respectively.  The observed synchrotron spectrum can be written as
{\small
\begin{eqnarray}
\label{espsyn}
\left[\epsilon^2_\gamma N(\epsilon_\gamma)\right]_{\rm \gamma, syn}= A_{\rm \gamma,syn}\cases{ 
(\frac{\epsilon_\gamma}{\epsilon^{\rm syn}_{\rm \gamma,m}})^\frac43    &  $\epsilon_\gamma < \epsilon^{\rm syn}_{\rm \gamma,m}$,\cr
 (\frac{\epsilon_\gamma}{\epsilon^{\rm syn}_{\rm \gamma,m}})^{-\frac{\alpha_e-3}{2}}  &  $\epsilon^{\rm syn}_{\rm \gamma,m} < \epsilon_\gamma < \epsilon^{\rm syn}_{\rm \gamma,c}$,\cr
(\frac{\epsilon^{\rm syn}_{\rm \gamma,c}}{\epsilon^{\rm syn}_{\rm \gamma,m}})^{-\frac{\alpha_e-3}{2}}    (\frac{\epsilon_\gamma}{\epsilon_{\rm \gamma,c}})^{-\frac{\alpha_e-2}{2}},           &  $\epsilon^{\rm syn}_{\rm \gamma,c} < \epsilon_\gamma < \epsilon^{\rm syn}_{\rm \gamma,max}$\,,\cr
}
\end{eqnarray}
}
\noindent where $A_{\rm \gamma,syn}$ is the proportionality constant of synchrotron spectrum. This constant can be estimated through the total number of radiating electrons in the volume of emitting region, ${\small n_e=N_e/V=4\pi N_e\,r_d^3/3}$, the maximum radiation power ${\small P_{\rm \nu,max}\simeq  \frac{dE_e/dt}{\epsilon_\gamma(\gamma_e)}}$ and the distance d$_z$ from the source.  Therefore, the proportionality constant can be written as
\bary
\label{Asyn}
 A_{\rm \gamma,syn}&=& \frac{P_{\rm \nu,max} n_e}{4\pi d_z^2}\,\epsilon^{\rm syn}_{\rm \gamma,m} \cr
&=&\frac{4\,\sigma_T}{9}\, d^2_z\,\delta^3_D\,U_B\,r_d^3\,N_e\,\gamma^2_{\rm e,min}\,.
\eary
\subsubsection{Compton scattering emission}
Fermi-accelerated electrons in the emitting region can upscatter synchrotron photons up to higher energies as 
\be\label{ic}
\epsilon^{\rm ssc}_{\gamma,{\rm (m,c,max)}}\simeq  \gamma^2_{e,{\rm (min,c,max)}} \epsilon^{\rm syn}_{\gamma,{\rm (m,c, max)}}\,.
\label{ic}
\ee
From  the electron Lorentz factors (eqs. \ref{gamma_m}, \ref{gamma_b}, \ref{gamma_max}) and the synchrotron break energies (eq. \ref{synrad}), we get that the Compton scattering break energies are given in the form  
\begin{eqnarray}\label{icrad}
\epsilon^{\rm ssc}_{\rm \gamma,m} &=& \frac{\sqrt{8\pi}\,q_e}{m_e}\,\delta_D\,U_B^{1/2}\,\gamma^4_{\rm e,min},\cr
\epsilon^{\rm ssc}_{\rm \gamma,c} &=&\frac{ 81\sqrt{2\pi}\,q_e\,m_e^3}{128\,\sigma_T^4}\,(1+Y)^{-4}\,\delta_D^5\, U_B^{-7/2}\, r_d^{-4},\cr
\epsilon^{\rm ssc}_{\rm \gamma, max} &=&\frac{9\,q_e^3}{2\sqrt{2\pi}\,m_e\,\sigma^2_T}\, \delta_D\,U_B^{-1/2}\,.
\end{eqnarray}
The Compton scattering spectrum  obtained  as a function of the synchrotron spectrum (eq. \ref{espsyn}) is
{\small
\begin{eqnarray} \label{espic}
\left[\epsilon^2_\gamma N(\epsilon_\gamma)\right]_{\rm \gamma, ssc}=A_{\rm \gamma,ssc} \cases{ 
(\frac{\epsilon_\gamma}{\epsilon^{\rm ssc}_{\rm \gamma,m}})^\frac43    &  $\epsilon_\gamma < \epsilon^{ssc}_{\gamma,m}$,\cr
 (\frac{\epsilon_\gamma}{\epsilon^{\rm ssc}_{\rm \gamma,m}})^{-\frac{\alpha_e-3}{2}}  &  $\epsilon^{\rm ssc}_{\rm \gamma,m} < \epsilon_\gamma < \epsilon^{\rm ssc}_{\rm \gamma,c}$,\cr
(\frac{\epsilon^{\rm ssc}_{\rm \gamma,c}}{\epsilon^{ssc}_{\gamma,m}})^{-\frac{\alpha_e-3}{2}}    (\frac{\epsilon_\gamma}{\epsilon^{ssc}_{\gamma,c}})^{-\frac{\alpha_e-2}{2}},           &  $\epsilon^{\rm ssc}_{\rm \gamma,c} < \epsilon_\gamma < \epsilon^{\rm ssc}_{\rm \gamma,max} $
}\cr
\end{eqnarray}
}
where $A_{\rm \gamma,ssc}=Y\,[\epsilon^2_\gamma N_\gamma(\epsilon_\gamma)]_{\rm max}^{\rm syn}$ is the proportionality constant of Compton scattering spectrum.
\subsection{Hadronic Model}
RGs have been proposed as a powerful accelerator of charged particles through the Fermi acceleration mechanism or/and magnetic reconnection \citep{2015arXiv150407592K}. We consider a proton population described as  a simple power law given by
\be\label{prot_esp}
\left(\frac{dN}{dE}\right)_p=A_p\,E_p^{-\alpha_p}\,,
\ee
with $A_p$ the proportionality constant and $\alpha_p$ the spectral power index of the proton population.  From eqs. (\ref{prot_esp})  we can compute that the proton density can be written as
\be
U_p=\frac{L_p}{4\pi\,\delta^2_D\,r^2_d}\,,
\ee
with the proton luminosity given by 
\be\label{lum}
L_p= 4\,\pi\,d^2_z A_p\,\int \,E_p\,E_p^{-\alpha_p}  dE_p\,.
\ee
Fermi-accelerated protons  lose their energies by electromagnetic channels and hadronic interactions.  Electromagnetic channels such as proton synchrotron radiation and inverse Compton will not be considered here, we will only assume that protons will be cooled down  by p$\gamma$ interactions at the emitting region of the inner jet. Charged ($\pi^+$) and neutral ($\pi^0$) pions are obtained from p$\gamma$ interaction through the following channels   
 \begin{eqnarray}
p\, \gamma &\longrightarrow&
\Delta^{+}\longrightarrow
\left\{
\begin{array}{lll}
p\,\pi^{0}\   &&   \mbox{fraction }2/3, \\
n\,  \pi^{+}      &&   \mbox{fraction }1/3\,.\nonumber
\end{array}\right. \\
\end{eqnarray}
After that neutral pion decays into photons, $\pi^0\rightarrow \gamma\gamma$,  carrying $20\% (\xi_{\pi^0}=0.2)$ of the proton's energy $E_p$.   The efficiency of the photo-pion production is \citep{1968PhRvL..21.1016S, PhysRevLett.78.2292}
{\small
\begin{equation}\label{eficiency}
f_{\pi^0} \simeq \frac {t_{\rm dyn}} {t^{-1}_{\pi^0}}  =\frac{r_d}{2\gamma^2_p}\int\,d\epsilon\,\sigma_\pi(\epsilon)\,\xi_{\pi^0}\,\epsilon\int dx\, x^{-2}\, \frac{dn_\gamma}{d\epsilon_\gamma} (\epsilon_\gamma=x)\,,
\end{equation}
}
where $dn_\gamma/d\epsilon_\gamma$ is the spectrum of seed photons,  $\sigma_\pi(\epsilon_\gamma)$ is the cross section of pion production and $\gamma_p$ is the proton Lorentz factor. Solving the integrals we obtain 
{\small
\bary
f_{\pi^0} \simeq \frac{\sigma_{\rm p\gamma}\,\Delta\epsilon_{\rm res}\,\xi_{\pi^0}\, L_{\rm \gamma,IC}}{4\pi\,\delta_D^2\,r_d\,\epsilon_{\rm pk,ic}\,\epsilon_{\rm res}}
\cases{
\left(\frac{\epsilon^{\pi^0}_{\gamma}}{\epsilon^{\pi^0}_{\gamma,c}}\right)^{\beta_h-1}       &  $\epsilon_{\gamma} < \epsilon^{\pi^0}_{\gamma,c}$\cr
\left(\frac{\epsilon^{\pi^0}_{\gamma}}{\epsilon^{\pi^0}_{\gamma,c}}\right)^{\beta_l-1}                                                                                                                                                                                                            &   $\epsilon^{\pi^0}_{\gamma,c} < \epsilon_{\gamma}$\,,\cr
}
\eary
}
where $\beta_h$ and $\beta_l$ are the high-energy and low-energy photon index, respectively,  ($L_{\rm \gamma, IC}$) is the observed luminosity around the second SSC peak, $\Delta\epsilon_{\rm res}$=0.2 GeV,  $\epsilon_{\rm res}\simeq$ 0.3 GeV, $\epsilon_{\rm pk,ic}$ is the energy of the second SSC peak and  $\epsilon^{\pi^0}_{\gamma,c}$ is the break photon-pion energy given by 
\be
\epsilon^{\pi^0}_{\gamma,c}\simeq 31.87\,{\rm GeV}\, \delta_D^2\, \left(\frac{\epsilon_{\rm pk,ic}}{ {\rm MeV}}\right)^{-1}\,.
\label{pgamma}
\ee
It is worth noting that the target photon density and the optical depth can be obtained through the equations
\be
n_{\gamma}\simeq\frac{L_{\rm \gamma,IC}}{4\pi r^2_d\,\epsilon_{\rm pk,ic}}\,,
\label{den}
\ee
and
\be
\tau_{\gamma}\simeq\frac{L_{\rm \gamma,IC}\,\sigma_T}{4\pi r_d\,\delta_D\epsilon_{\rm pk,ic}}\,.
\ee
respectively.  Taking into account that  photons released  in the range $\epsilon_\gamma$ to $\epsilon_\gamma + d\epsilon_\gamma$ by protons in the range   $E_p$ and $E_p + dE_p$ are $f_{\pi^0}\,E_p\,(dN/dE)_p\,dE_p=\epsilon_{\pi^0,\gamma}\,(dN/d\epsilon)_{\pi^0,\gamma}\,d\epsilon_{\pi^0,\gamma}$, then photo-pion spectrum is given by
{\small
\bary
\label{pgammam}
\left[\epsilon^2_\gamma N(\epsilon_\gamma)\right]_{\rm \gamma, \pi^0}= A_{\rm p\gamma}  \left(\frac{\epsilon^{\pi^0}_{\gamma,c}}{\epsilon_{0}}\right)^{-\alpha_p+2} \cases{
\left(\frac{\epsilon^{\pi^0}_{\gamma}}{\epsilon^{\pi^0}_{\gamma,c}}\right)^{\beta_h+1-\alpha_p}           &  $ \epsilon_{\gamma} < \epsilon^{\pi^0}_{\gamma,c}$\cr
\left(\frac{\epsilon^{\pi^0}_{\gamma}}{\epsilon^{\pi^0}_{\gamma,c}}\right)^{\beta_l+1-\alpha_p}                                                                                     &   $\epsilon^{\pi^0}_{\gamma,c} < \epsilon_{\gamma}$\,,\cr
}
\eary
}
\noindent where the proportionality constant  $A_{\rm p\gamma}$  is in the form
\be\label{Apg}
A_{\rm p\gamma}= \frac{L_{\rm \gamma,IC}\,\sigma_{\rm p\gamma}\,\Delta\epsilon_{\rm res}\,\epsilon^2_0\,\left(\frac{2}{\xi_{\pi^0}}\right)^{1-\alpha_p}}{4\pi\,\delta_D^2\,r_d\,\epsilon_{\rm pk,ic}\,\epsilon_{\rm res}}\,A_p\,.
\ee
The value of $A_p$ is determined through the TeV $\gamma$-ray flux (eq. \ref{Apg}).
\section{High energy neutrino expectation}
Photo hadronic interactions in the emitting region (see subsection 3.2) also generate neutrinos through the charged pion decay products ($\pi^{\pm}\rightarrow \mu^\pm+  \nu_{\mu}/\bar{\nu}_{\mu} \rightarrow  e^{\pm}+\nu_{\mu}/\bar{\nu}_{\mu}+\bar{\nu}_{\mu}/\nu_{\mu}+\nu_{e}/\bar{\nu}_{e}$).  Taking into account the distances of RGs, the neutrino flux ratio (1 : 2 : 0 ) created on the source will arrive on the standard ratio (1 : 1 : 1 ) \citep{2008PhR...458..173B}.  The neutrino spectrum produced by the photo hadronic interactions is
{\small
\bary
\label{pgammam}\label{espneu1}
\left[\en^2 N(\epsilon_\nu)\right]_\nu= A_\nu \epsilon^2_0 \cases{
\left(\frac{\en}{\epsilon_0 }\right)^{\beta_h}                           &  $ \en < \enc$\cr
\left(\frac{\en}{\epsilon_0 }\right)^{\beta_l+1-\alpha_{\nu}}     &   $\enc < \en$\,,\cr
}
\eary
}
where the factor, A$_{\nu}$,  normalized through the TeV $\gamma$-ray flux is \citep[see, Julia Becker] [and reference therein]{2007Ap&SS.309..407H}
\be
A_\nu=A_{\rm p\gamma}\,\epsilon_0^{-2}\, 2^{-\alpha_p}\,.
\ee
The previous equation was obtained from solving the integral terms {\small $\int \frac{dN_{\nu}}{d\en}\,\en\,d\en=\frac14\int \frac{dN_\gamma}{dE_\gamma}\,E_\gamma\,dE_\gamma$}, considering that the spectral indices for neutrino and that $\gamma$-ray spectra are similar  $\alpha_p\simeq \alpha_\nu$ \citep{2008PhR...458..173B} and each neutrino (photon) bring 5\% (10\%) of the initial proton energy.
The neutrino flux is detected when it interacts inside the instrumented volume. Considering the probability of interaction for a neutrino with energy $E_\nu$ in an effective volume ($V_{eff}$) with density ($\rho_{ice}$), the number of expected neutrino events after a period of time $T$ is  
{\small
\be
N_{\rm ev} \approx\,T \rho_{\rm ice}\,N_A\,\epsilon_0\int_{E_{\rm \nu,th}} V_{\rm eff}(E_\nu) \sigma_{\rm \nu N}(E_\nu) \,A_\nu\left(\frac{E_{\nu}}{\epsilon_0}\right)^{-\alpha_p}\, dE_\nu\,,
\label{numneu1}
\ee
}
where $N_A$ is the Avogadro number, $\sigma_{\nu N}(E_\nu)$ is the the charged current cross section and $E_{\nu,th}$ is the  energy threshold.  It is worth noting that the effective volume $V_{eff}$ is obtained for a hypothetical Km$^{3}$ neutrino telescope  through the Monte Carlo simulation.
\section{Ultra-high-energy cosmic rays}
TeV $\gamma$-ray observations from low-redshift AGN have been proposed as good candidates for studying UHECRs \citep{2012ApJ...749...63M, 2009NJPh...11f5016D, 2012ApJ...745..196R}.   We consider that the proton spectrum is extended up to UHEs and then calculate the  number of events expected in PAO, TA and HiRes experiments.
\subsection{Hillas Condition}
By considering that super massive BHs have the power to accelerate particles  up to UHEs through Fermi processes,  protons accelerated in the emitting region are confined by the Hillas condition \citep{1984ARA&A..22..425H}. Although this requirement is a necessary condition and acceleration of UHECRs in AGN jets \citep{2012ApJ...749...63M,2012ApJ...745..196R,2010ApJ...719..459J},  it is far from trivial (see e.g., \cite{2009JCAP...11..009L} for a more detailed energetic limits). The Hillas criterion is defined as the maximum proton energy achieved in a region with radius $r_d$ and magnetic field $B$. It can be written as   
\be\label{Emax}
E_{\rm p,max}\simeq \frac{Zq_e}{\phi}\,B\,r_d\,\Gamma\,.
\ee
Here $Z$ is the atomic number, $\phi\simeq$ 1 is the acceleration efficiency factor and $\Gamma$ is the bulk Lorentz factor given by  
\be
\Gamma=\frac{1\pm\sqrt{1-(1-\cos^2\theta)(1+\delta^2_D\cos^2\theta)}}{\delta_D(1-\cos^2\theta)}\,,
\ee
where $\theta$ is the viewing angle.
\subsection{Deflections}
The magnetic fields play important roles on cosmic rays.  UHECRs traveling from source to Earth are randomly deviated by galactic (B$_G$) and extragalactic (B$_{EG}$) magnetic fields. By considering that magnetic fields are quasi-constant and homogeneous,  the deflection angle due to the B$_G$ and  B$_{EG}$  \citep{1997ApJ...479..290S} are
\be\label{thet_G}
\psi_{\rm G}\simeq 3.8^{\circ}\left(\frac{E_{p,th}}{57 EeV}\right)^{-1} \int^{L_G}_0  | \frac{dl}{{\rm kpc}}\times \frac{B_G}{4\,{\rm \mu G}} |\,,
\ee
and
\be\label{thet_EG}
\psi_{\rm EG}\simeq 4^{\circ}\left(\frac{E_{p,th}}{57 EeV}\right)^{-1} \,\left(\frac{L_{\rm EG}}{100\, {\rm Mpc}}\right)^{1/2}\,\left(\frac{l_c}{1\, {\rm Mpc}}\right)^{1/2}\,\left( \frac{B_{\rm EG}}{1\,{\rm nG}} \right)\,,   
\ee
respectively, where L$_{\rm G}$ corresponds to the distance of our Galaxy (20 kpc), $l_c$ is the coherence length and $E_{\rm p,th}$ is the threshold proton energy.   Due to the strength of extragalactic ($B_{\rm EG}\simeq$ 1 nG) and galactic ($B_{\rm G}\simeq$ 4 $\mu$G) magnetic fields,   UHECRs are deflected; firstly, $\psi_{\rm EG}\simeq 4^{\circ}$ and after $\psi_G\simeq 3.8^{\circ}$, between the true direction to the source and the observed arrival direction, respectively.  Estimation of the deflection angles could associate the transient UHECR sources with the HE neutrino and $\gamma$-ray fluxes. Taking into account of extragalactic and galactic magnetic fields, it is reasonable to correlate UHECRs lying within  $\sim5^\circ$ of a source. 
\subsection{Expected number of events}
\paragraph{Telescope Array observatory}.   Located in Millard Country (Utah), TA experiment was designed to study UHECRs with energies above 57 EeV.  With an area of $\sim$ 700 km$^2$,  it is made of  a scintillator surface detector (SD) array and three fluorescence detector (FD) stations \citep{2012NIMPA.689...87A}.   The TA  exposure is given by $\Xi\,t_{op}\, \omega(\delta_s)/\Omega$, where $\Xi\,t_{op}=(5)\,7\times10^2\,\rm km^2\,yr$, $t_{op} $ is the total operational time (from 2008 May 11 and 2013 May 4),  $\omega(\delta_s)$ is an exposure correction factor for the declination of the source \citep{2001APh....14..271S} and $\Omega\simeq\pi$. 
\paragraph{Pierre Auger observatory}.  The PAO, located in the Mendoza Province of Argentina at latitude $\simeq 36^\circ$S, was designed to determine the arrival directions and energies of UHECRs using four fluorescence telescope arrays and 1600 surface detectors spaced 1.5 km apart. The large exposure of its ground array, combined with accurate energy has provided an opportunity to explore the spatial correlation between cosmic rays and their sources in the sky.   The PAO  exposure is given by $\Xi\,t_{op}\, \omega(\delta_s)/\Omega_{60}$, where $\Xi\,t_{op}=9\times10^3\,\rm km^2\,yr$, $t_{op} $ is the total operational time (from 1 January 2004 until 31 August 2007),  $\omega(\delta_s)$ is an exposure correction factor and $\Omega_{60}\simeq\pi$ is the Auger acceptance solid angle \citep{2007Sci...318..938P, 2008APh....29..188P}.  
\paragraph{The High Resolution Fly's eye experiment}.  The HiRes experiment, located atop two hills 12.6 km apart in the west-central Utah desert, consisted of two detectors that observed cosmic ray showers via the fluorescence light.  Both detectors, called HiRes-I and HiRes-II, consisted of 21 and 42 telescopes, respectively, each one composed of a spherical mirror of 3.8 m$^2$. The HiRes exposure  is (3.2 - 3.4) $\times 10^3$ km$^2$ year sr.   HiRes experiment  measured the flux of UHECRs using the stereoscopic air fluorescence technique  over a period of nine years (1997 - 2006) \citep{2005ApJ...622..910A, 2009APh....32...53H}.\\
The expected number of UHECRs above an energy $E_{p,th}$ yields
\be
N_{\rm \tiny UHECR}=  ({\rm  Expos.})\times \,N_p, 
\label{num}
\ee
where $N_p$ is calculated from the proton spectrum extended up to energies higher than $E_{p,th}$ (eq. \ref{prot_esp}).   The expected number can be written as
\bary
N_{\rm \tiny UHECR}=\frac{\Xi\,t_{\rm op}\, \omega(\delta_s)}{(\alpha_p-1)\Omega}\,A_p  \int_{E_{\rm p,th}} E_{p}^{-\alpha_p} dE_p\,,
\label{nUHE1}
\eary
where the value of $A_p$ is normalized with the TeV $\gamma$-ray fluxes (eq. \ref{Apg}).
\section{Results and Discussion}
\begin{figure}
\centering
\includegraphics[width=1\textwidth]{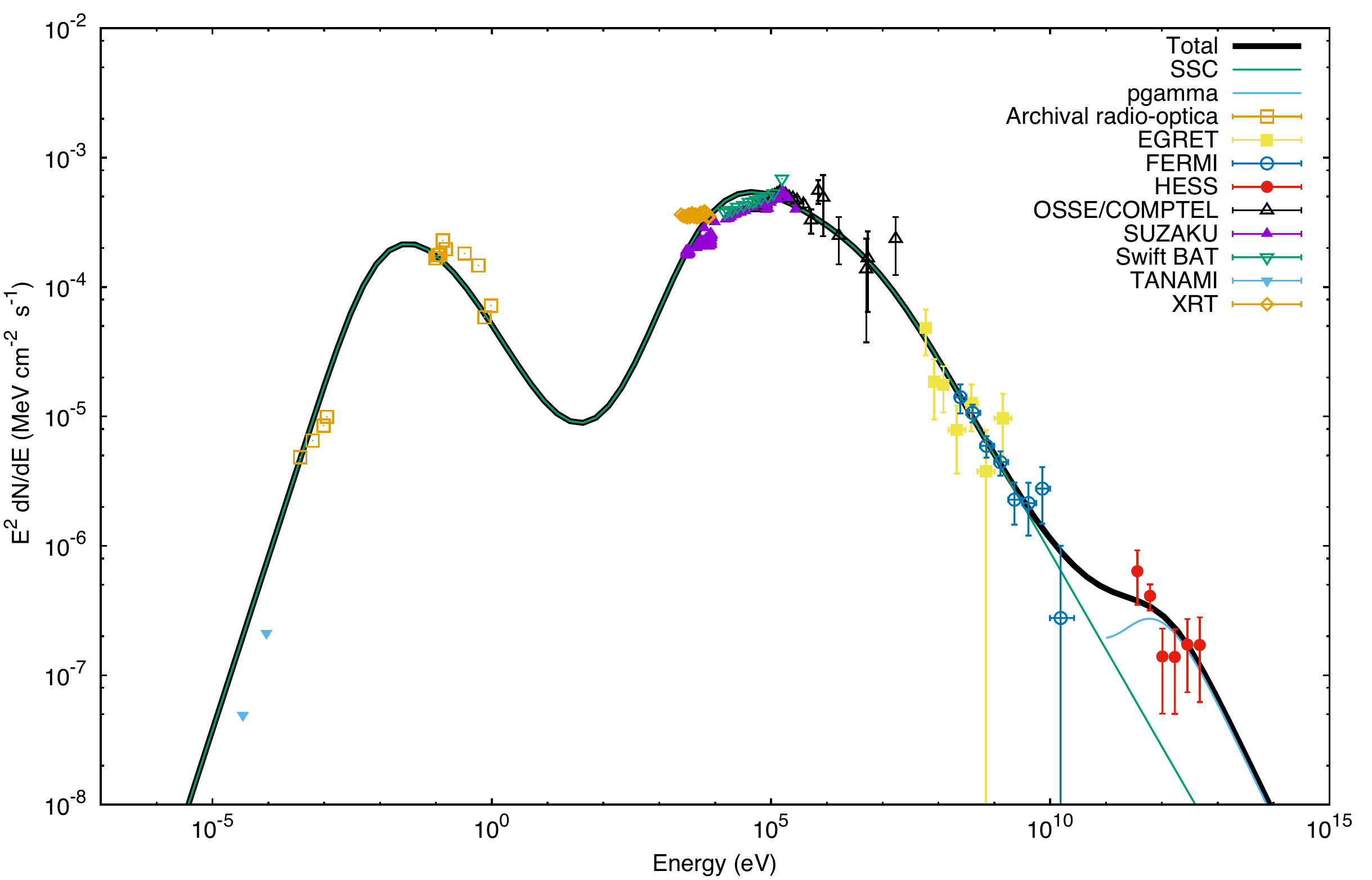}
\caption{SED of the Cen A core with our model fit (black line).  The green line is a SSC fit to the TANAMI VLBI \citep{2010arXiv1001.0059O}, archival \citep{2000ApJ...528..276M}, Suzaku  \citep{2007ApJ...665..209M}, Swift-XRT  \citep{2005A&A...440..775K}, Swift-BAT \citep{2009ApJ...690..367A},  COMPTEL \citep{1998A&A...330...97S}, EGRET \citep{1999ApJS..123...79H, 1999APh....11..221S} and Fermi-LAT \citep{2010ApJ...719.1433A} data.  The blue line is a fit to the HESS \citep{2009ApJ...695L..40A} data using the $\pi^0$ decay product from p$\gamma$ interactions.} \label{cenA}
\end{figure}

\begin{figure}
\centering
\includegraphics[width=0.8\textwidth]{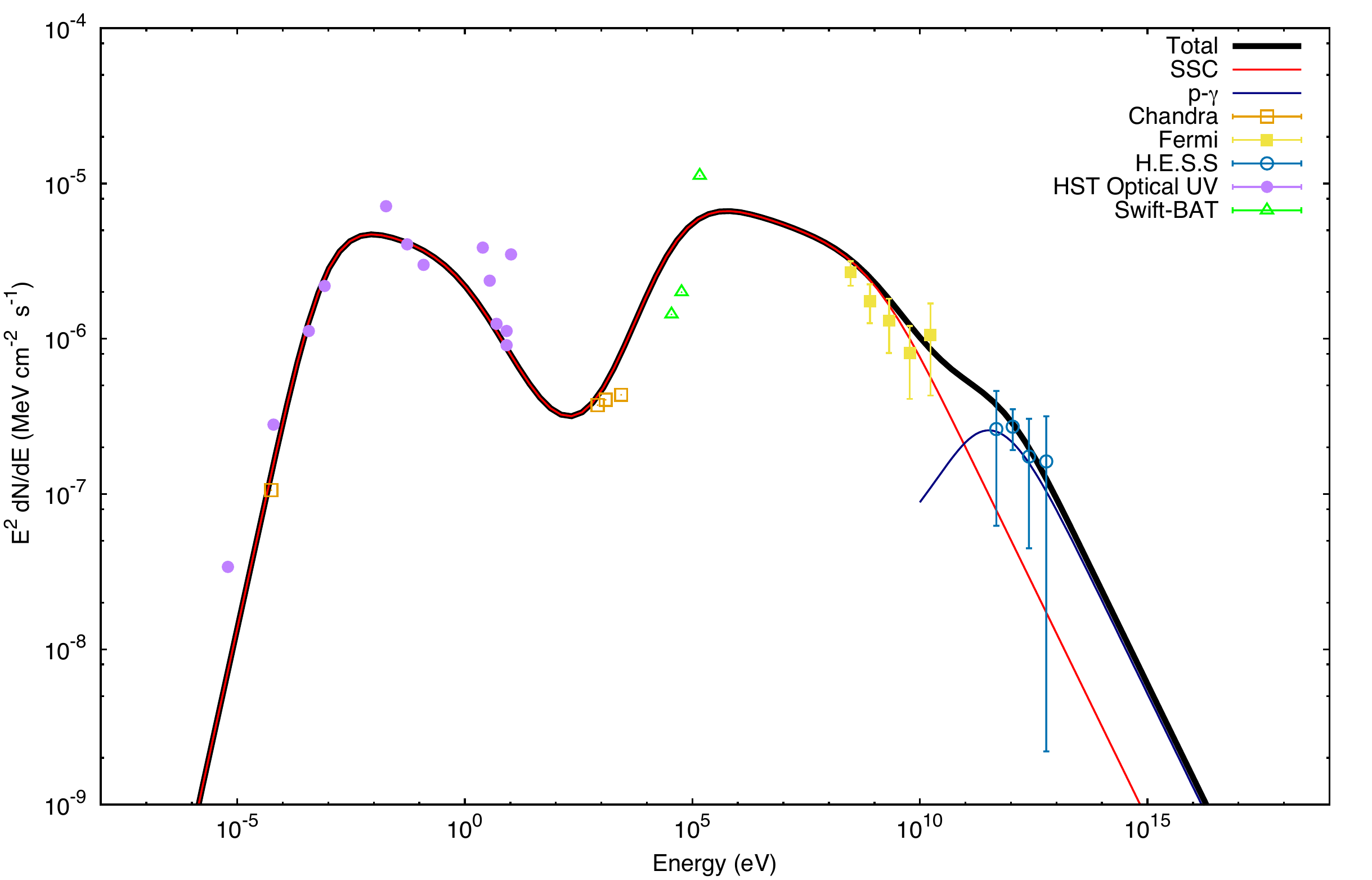}
\caption{SED of the M87 core with our model fit (black line).  The red line is a SSC fit to the VLA - HST \citep{1991AJ....101.1632B, 1996A&A...309..375D, 1996ApJ...473..254S, 2007ApJ...668L..27K, 2001ApJ...561L..51P, 2007ApJ...655..781S, 2008ApJ...689..775T}, Swift-BAT \citep{2009ApJ...690..367A},  Chandra  \citep{2002ApJ...564..683M}.  The blue line is a fit to the HESS \citep{2006Sci...314.1424A} data using the $\pi^0$ decay product from p$\gamma$ interactions.}. \label{m87}
\end{figure}

\begin{figure}
\centering
\includegraphics[width=0.8\textwidth]{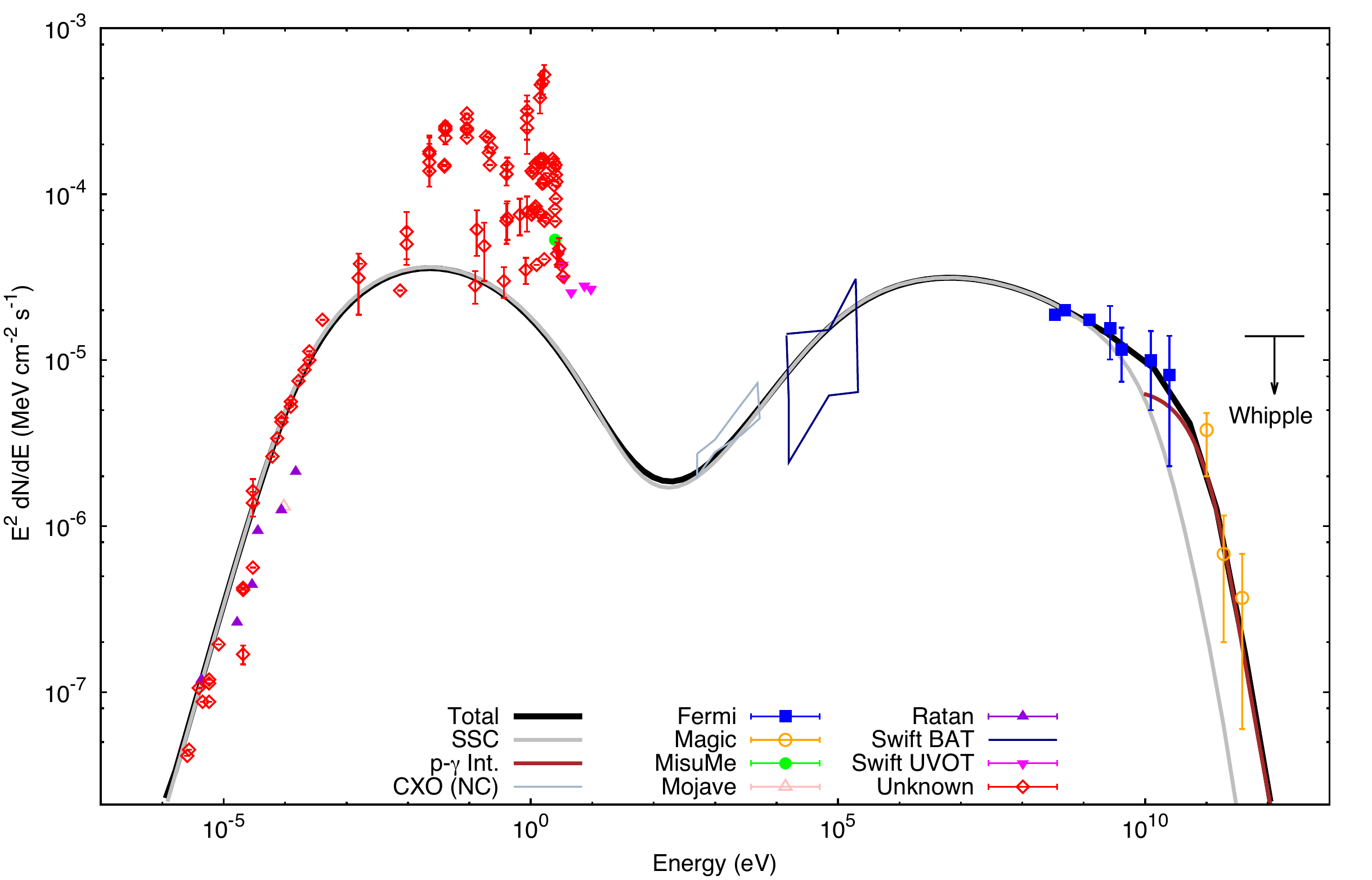}
\caption{SED of the NGC1275 core with our model fit (black line). The gray line is a SSC fit to the  RATAN \citep{1999A&AS..139..545K}, MOVAVE \citep{2009AJ....137.3718L} , MITSuME \citep{2005NCimC..28..755K}, Swift-BAT \citep{2009ApJ...690..367A}, Swift-UVOT \citep{2005SSRv..120...95R} and Fermi-LAT \citep{2009ApJ...699...31A} data.  The red line is a fit to the Magic \citep{2010ApJ...710..634A} data using the $\pi^0$ decay product from p$\gamma$ interactions.  Whipple upper limit is included \citep{2006ApJ...644..148P}.}. \label{ngc1275}
\end{figure}

\begin{figure}
\centering
\includegraphics[width=0.8\textwidth]{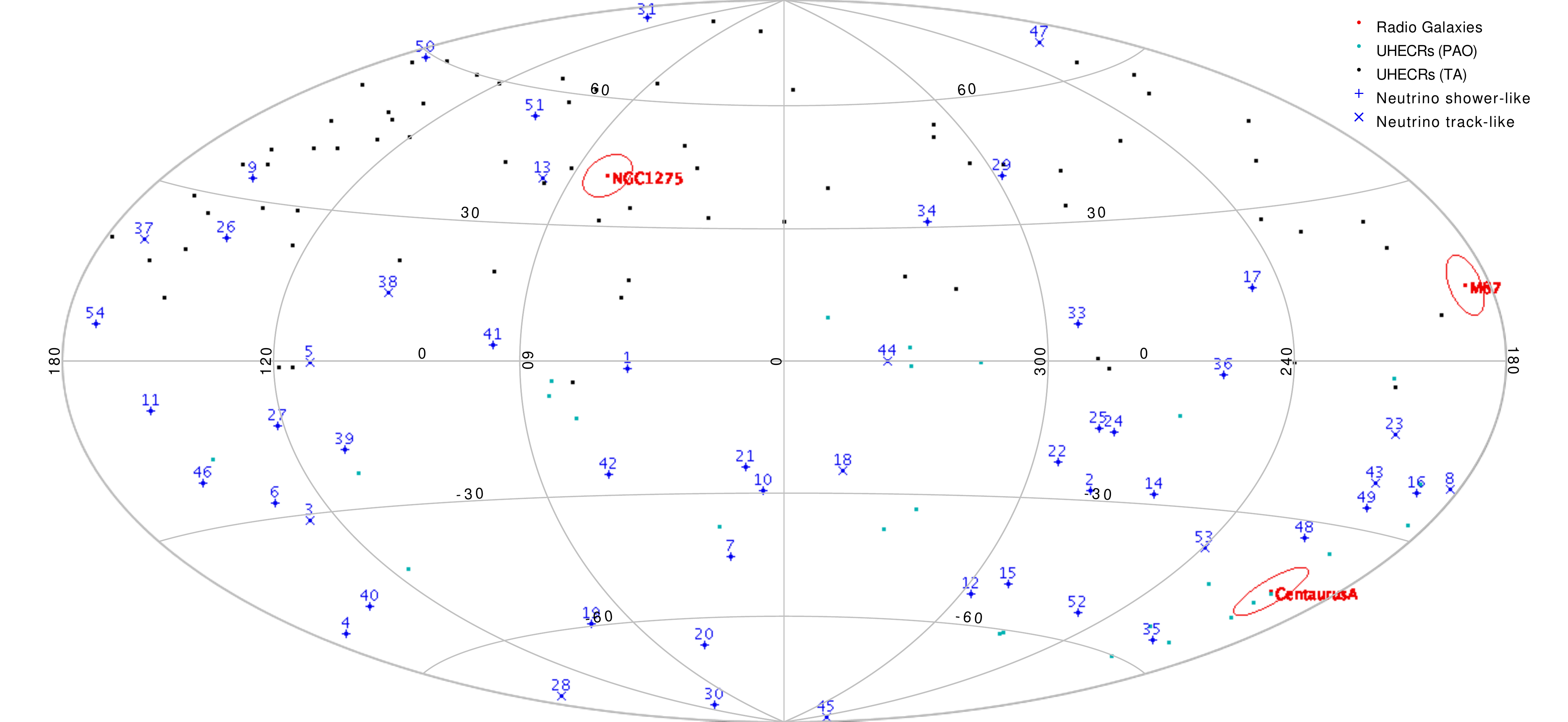}\\
\caption{Skymap in equatorial coordinates of UHECRs, neutrino events and ll AGN.  Black and green points are the UHECRs reported by TA and PAO  Collaborations, respectively. In blue are the neutrino events ({\bf X} for  like-track and {\bf {\Large +}} for like-shower) reported by IceCube Collaboration.  For the closest RGs we reported the Cen A, M87 and NGC 1275 positions with red circles of $5^{\circ}$ around them.}\label{Skymap}
\end{figure}

\begin{figure}
\centering
\includegraphics[width=0.7\textwidth]{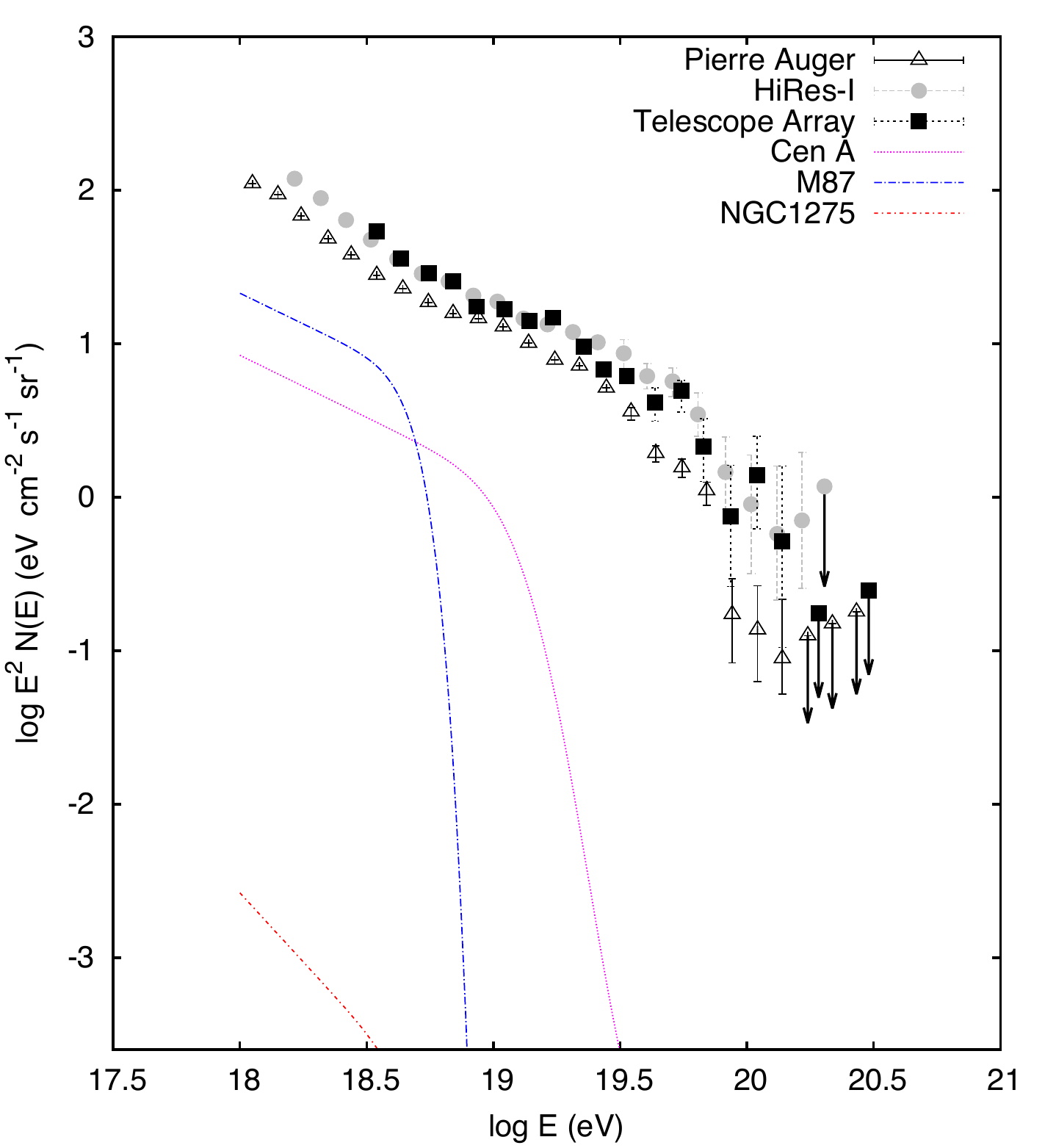}
\caption{The UHECR spectra collected with PAO \citep{2011arXiv1107.4809T}, HiRes \citep{2009APh....32...53H} and TA \citep{2013ApJ...768L...1A} experiment  are overlapped with UHE proton fluxes of RGs (Cen A, M87 and NGC1275) resulting from extrapolating the proton fluxes used to describe the TeV $\gamma$-ray spectra.}. \label{uhecr_flux}
\end{figure}

\begin{figure}
\centering
\includegraphics[width=0.7\textwidth]{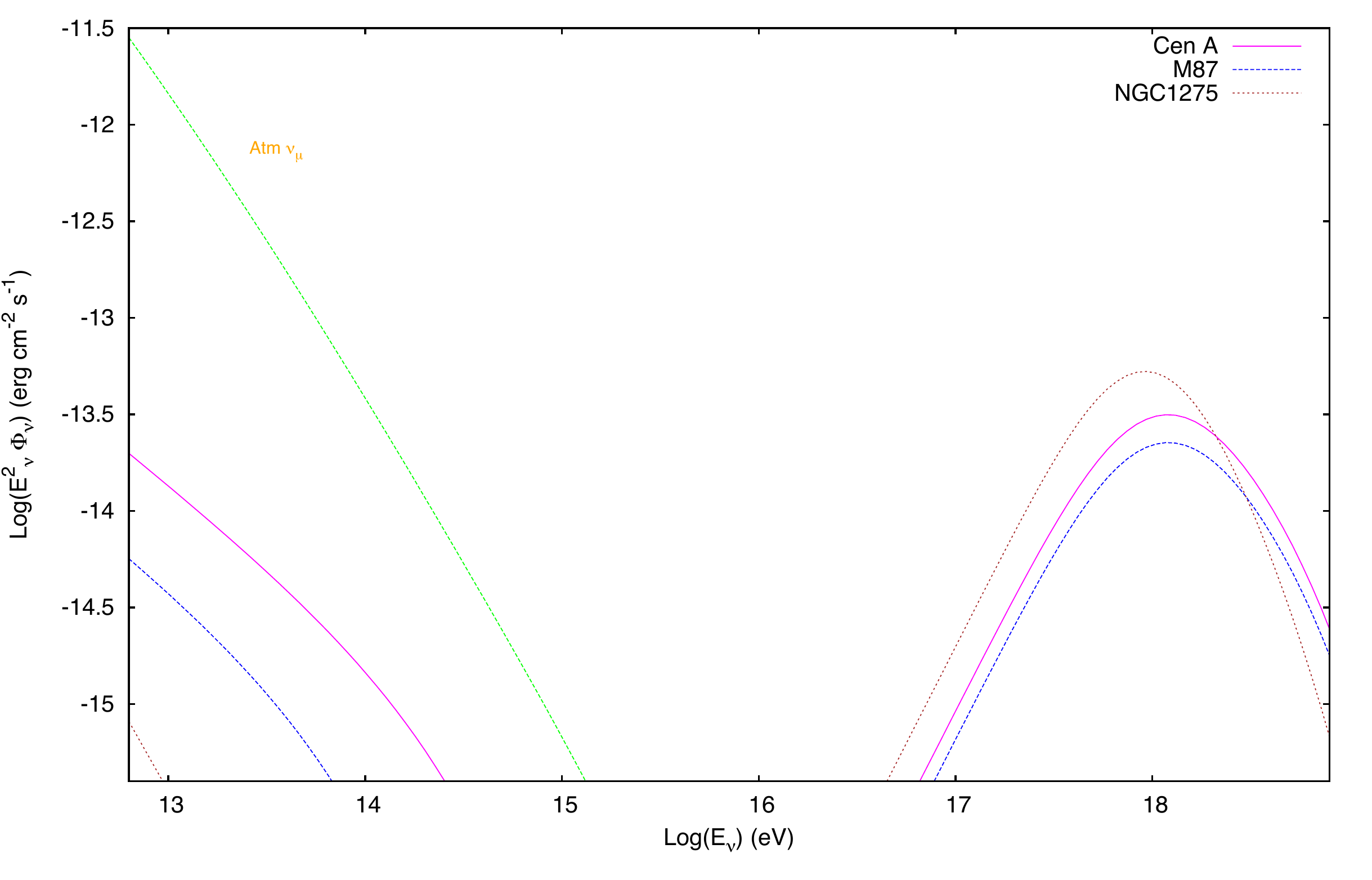}
\caption{Neutrino spectra of RGs (Cen A, M87 and NGC1275) obtained as $\pi^\pm$ decay products from p$\gamma$ interactions of Fermi-accelerated protons with the seed photons around the first and second SED peaks. The atmospheric muon neutrino background is also shown.} \label{neutr_flux}
\end{figure}

\begin{figure}
\centering
\includegraphics[width=0.9\textwidth]{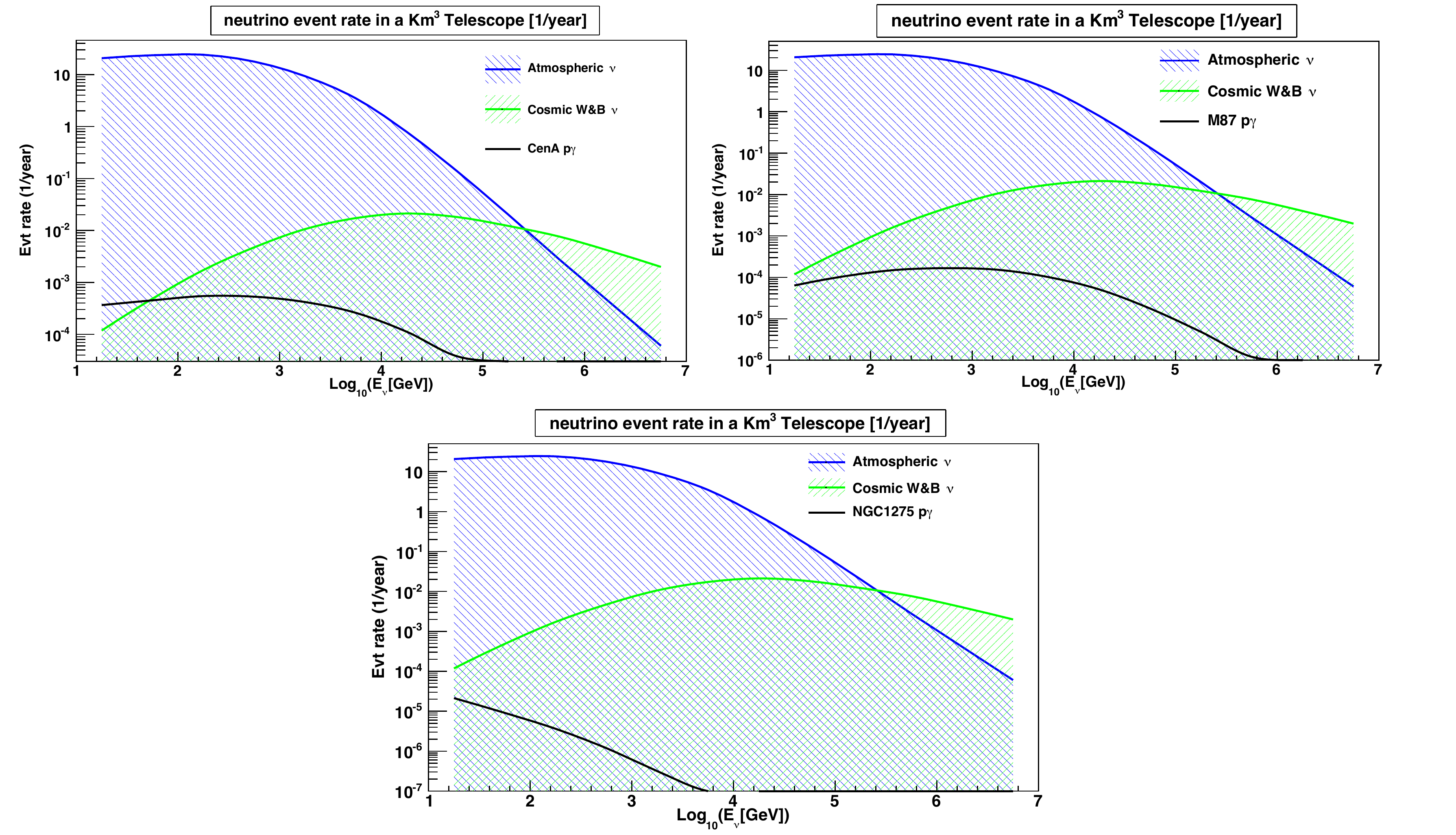}
\caption{Neutrino signal to noise ratio for a Km$^{3}$ neutrino telescope for Cen A (left), M87 (right) and NGC1275 (below). The black lines represent the neutrino signal produced by p$\gamma$ interactions respectively. The blue and the green areas represent the atmospheric and diffuse neutrino ``background'' within a region of 1$^{\circ}$ around the source position.} \label{neut_event}
\end{figure}

We have presented a lepto-hadronic model to describe the broadband SED of the closest RGs, supposing that electrons and protons are co-accelerated at the emitting region of the jet.  In the leptonic model, we have required the SSC model to explain the spectrum up to dozens of GeV and in the hadronic scenario, we have evoked the p$\gamma$ interactions occurring close to the core of the RGs to interpret the TeV $\gamma$-ray fluxes.  The SSC model depends basically on magnetic field ($B$), electron density ($N_e$), size of emitting region ($r_d$) and Doppler factor ($\delta_D$).  To reproduce the electromagnetic spectrum up to  dozens of GeV, we have used an electron distribution described by a broken power law (eq. \ref{espele}) with the minimum, cooling and maximum Lorentz factors given by eqs.  (\ref{gamma_m}), (\ref{gamma_b}), (\ref{gamma_max}), respectively.  The minimum Lorentz factor is obtained through electron density and  electron energy density,  the cooling Lorentz factor is computed through the synchrotron and dynamical time scales, and the maximum Lorentz factor is calculated considering the synchrotron and the acceleration time scales.  Considering the break Lorentz factors, the synchrotron and Compton scattering break energies are obtained (eqs. \ref{synrad} and \ref{icrad}).  The synchrotron (eq. \ref{espsyn})  and Compton scattering (eq. \ref{espic}) spectra are estimated through the density and distribution of radiating electrons confined inside emitting region.   In the p$\gamma$ interaction model,  we have considered  Fermi-accelerated  protons described by  a simple power law (eq. \ref{prot_esp}) which are accelerated close to the core and furthermore interact with the photon population at the second-peak SED photons.  The spectrum generated by this hadronic process (eq. \ref{pgammam}) depends on the proton luminosity (through $A_p$),  the observed luminosity around the second SSC peak ($L_{\rm \gamma, IC}$), the energy of the second SSC peak ($\epsilon_{\rm pk,ic}$), the size of emitting region, the Doppler factor,  and the high-energy and low-energy photon index, respectively.  The efficiency of the photo-production (eq. \ref{eficiency}) is calculated through the photo-pion cooling and dynamical time scales for $\beta_h\sim2$ and $\beta_l\sim1$.   Evoking these interactions,  we have interpreted naturally the TeV $\gamma$-ray spectra as $\pi^0$ decay products.\\
We have required the data used by the Fermi collaboration for Cen A  \citep{2010ApJ...719.1433A}, M87 \citep{2009ApJ...707...55A} and NGC1275 \citep{2009ApJ...699...31A}. In the case of NGC1275,  the TeV $\gamma$-ray data have been added \citep{2012A&A...539L...2A}.  Using the method of Chi-square $ \chi^2$ minimization as implemented in the ROOT software package \citep{1997NIMPA.389...81B}, we get the break energies, spectral indexes and normalization of the SSC and photo-pion spectrum  as reported in Table 1 and Figures \ref{cenA},  \ref{m87} and \ref{ngc1275} for Cen A , M87 and NGC1275, respectively.   For the sake of simplicity, the process was integrated into a python script that called the ROOT routines via the pyroot module. The whole data of the closest RGs were read from a file and written into an array which was then fitted with pyROOT.  The fitted data with the PyROOT module and from eqs. (\ref{espsyn}), (\ref{espic}) and (\ref{pgammam})  were plotted with a smooth sbezier curve through the gnuplot software \footnote{gnuplot.sourceforge.net}.   The sbezier option first renders the data monotonic (unique) and then applies the Bezier algorithm\footnote{Bezier curves are widely used in computer graphics to model smooth curves using the Casteljau algorithm which is a method to split a single Bezier curve into two Bezier curves at an arbitrary parameter value.  http://web.mit.edu/hyperbook/Patrikalakis-Maekawa-Cho/node13.html}.
%
%
%
\begin{center}\renewcommand{\arraystretch}{1.3}\addtolength{\tabcolsep}{4pt}
\begin{center}
\scriptsize{\textbf{Table 1.  Values obtained after fitting the whole spectrum of the RGs with our lepto-hadronic model.}}\\
\end{center}
\begin{tabular}{ l c c c c}
 \hline \hline
 \scriptsize{Parameter} &\scriptsize{Symbol} & \scriptsize{Cen A}& \scriptsize{M87} & \scriptsize{NGC1275}\\
 \hline
\hline
\multicolumn{2}{c}{Leptonic model} \\
\cline{1-2}
\scriptsize{$ A_{\rm syn,\gamma}\,\,  ({\rm MeV\,cm^{-2}\,s^{-1}})$}  &\scriptsize{[0]}  &   \scriptsize{$(0.25\pm 0.05)\times 10^{-4}$}&  \scriptsize{$(0.66\pm 0.08)\times 10^{-4}$} &   \scriptsize{$(1.25\pm 0.26)\times 10^{-5}$} \\
\scriptsize{$\alpha_e$} & \scriptsize{[1]}   & \scriptsize{3.50$\pm$ 0.02} & \scriptsize{3.21$\pm$ 0.02} &  \scriptsize{2.81$\pm$ 0.05} \\
 \scriptsize{$ \epsilon^{syn}_{\rm \gamma,m}\,\, ({\rm eV})$}  & \scriptsize{[2]}  & \scriptsize{ $0.046\pm 0.002$}& \scriptsize{$(2.26\pm 0.25)\times 10^{-2}$}  & \scriptsize{$(1.25\pm 0.05)\times 10^{-2}$} \\
\scriptsize{$\epsilon^{syn}_{\rm \gamma,c}\,\, ({\rm eV})$}  & \scriptsize{[3]}   & \scriptsize{0.15$\pm$ 0.02} & \scriptsize{$0.15\pm 0.02$} &  \scriptsize{ $0.33\pm 0.02$} \\
\hline
\scriptsize{$ A_{\rm ssc,\gamma}\,\,  ({\rm MeV\,cm^{-2}\,s^{-1}})$}  &\scriptsize{[4]}  &   \scriptsize{$(6.65\pm 0.67) \times 10^{-4}$} &  \scriptsize{ $(1.31\pm 0.75) \times 10^{-5}$} &  \scriptsize{ $(3.99\pm 0.86) \times 10^{-4}$} \\
\scriptsize{$ \epsilon^{\rm ssc}_{\rm \gamma,m}\,\, ({\rm keV}) $}  & \scriptsize{[5]}  & \scriptsize{ $67.1\pm 1.51$} &   \scriptsize{ $11.2\pm 0.9$}  &  \scriptsize{$3.15\pm 0.16$}  \\
\scriptsize{$\epsilon^{\rm ssc}_{\rm \gamma,c}\,\, ({\rm MeV})$}  & \scriptsize{[6]}   & \scriptsize{$0.53\pm 0.02$} & \scriptsize{$0.41\pm 0.15$} &   \scriptsize{$2.1\pm 0.1$}  \\
\hline
\multicolumn{2}{c}{Hadronic model} \\
\cline{1-2}
\scriptsize{$ A_{\rm p\gamma} \,\,  ({\rm MeV\,cm^{-2}\,s^{-1}})$}   & \scriptsize{[7]}   &   \scriptsize{$(3.98\pm 0.18)\times 10^{-7}$}& \scriptsize{$(1.17\pm 0.24)\times 10^{-7}$} &  \scriptsize{$(3.69\pm 0.91)\times 10^{-7}$} \\
\scriptsize{$\alpha_p$} & \scriptsize{[8]}   & \scriptsize{2.81$\pm$ 0.05}&  \scriptsize{2.80$\pm$ 0.02}  &  \scriptsize{ $3.81\pm 0.64$} \\
 \hline
\end{tabular}
\end{center}
%
%
From eqs. (\ref{synrad}), (\ref{Asyn}), (\ref{icrad}) and the values reported in Table 1, we have obtained the values of magnetic field, electron density, size of emitting region and Doppler factor that describe the broadband SED of Cen A, M87 and NGC1275. For this fit, we have considered the effect of the electromagnetic background light (EBL) absorption \citep{2008A&A...487..837F} and adopted the typical values reported in the literature such as viewing angles,  observed luminosities, distances of the closest RGs, minimum Lorentz factors and energy normalizations \citep{2007ApJ...654..186F,1991ApJ...373L...1T, 1992ApJS...83...29S, 2010ApJ...719.1433A, 1996ApJ...467..597B, 2006PASJ...58..261A, 2009ApJ...695L..40A, 2006Sci...314.1424A, 2012A&A...539L...2A, 2009ApJ...707...55A, 2009ApJ...699...31A, 1998A&A...330...97S, 2009ApJ...690..367A}.  Other quantities such as bulk Lorentz factors, proton and electron luminosities, magnetic field, proton and electron densities, etc, are derived from these parameters.  Table 2 shows all the parameter values obtained, used and derived in and from the fit.\\
The maximum proton Lorentz factors were estimated with the maximum electron Lorentz factors as {\small $\gamma_{p,max}=\frac{m_p}{m_e}\gamma_{e,max}$} \citep{2014A&A...562A..12P}.  Requiring that electron and proton number densities are similar ($N_e\simeq N_p$), we have calculated the minimum  proton Lorentz factors $\gamma_{p,min}$. These values could increase/decrease considering that proton number density is smaller/higher than the electron number density.  For instance,  if we assume that ($N_p\simeq  b\, N_e$) with b=10 (0.1) the minimum proton Lorentz factors are  $\gamma_{p,min}= 9.1\times10^3$ ($1.6\times10^6$), $3.2\times10^5$ ($6.7\times10^7$) and $7.1\times10^5$ ($9.2\times10^6$) for Cen A, M87 and NGC1275, respectively.   Due to the fact that the minimum Lorentz factor cannot be determined just assuming the standard scenario of injection and acceleration, we have used the condition that electron and proton number densities are similar.  Otherwise,  for $\gamma_{p,min}=1$ the proton luminosities becomes $>5\times 10^{47}\, {\rm erg/s}$.  Considering  the values of magnetic field, electron and proton energy densities, and their rates;  $\lambda_{e,B}=\frac{U_e}{U_B}$ ($\lambda_{p,B}=\frac{U_p}{U_B}$)= 2.92 (4.90), 24.1 (26.9) and 117.81 (382.56) for Cen A, M87 and NGC1275, respectively, we can see that energy densities could be related through principle of equipartition.\\ 
%
%
%
\begin{center}\renewcommand{\arraystretch}{1.1}\addtolength{\tabcolsep}{4pt}
\vspace{4cm}
\begin{center}
\scriptsize{\textbf{Table 2. Parameters obtained, derived and used of lepton-hadronic model to fit  the spectrum of Cen A, M87 and NGC1275.}}\\
\end{center}
\begin{tabular}{ l c c c c c}
\hline
\hline
\normalsize{} & \normalsize{Cen A}& \normalsize{M87}&\normalsize{NGC1275}  \\
\hline
\hline
\multicolumn{2}{c}{Obtained quantities} \\
\cline{1-2}
\scriptsize{$\delta_d$} & \scriptsize{1.0} & \scriptsize{2.8}&  \scriptsize{2.2} \\
\scriptsize{$B$ (G)} & \scriptsize{3.6}&  \scriptsize{1.61}&  \scriptsize{2.01} \\
\scriptsize{$r_d$ (cm)} & \scriptsize{$5.2 \times 10^{15}\,$} & \scriptsize{$2.1 \times 10^{15}\,$}   & \scriptsize{$2.26 \times 10^{15}\,$}  \\
\scriptsize{$N_e$ (cm$^{-3}$)} & \scriptsize{$1.1\times 10^3$} &   \scriptsize{$2.3\times 10^3$}&  \scriptsize{$3.11\times 10^4$} \\\hline
\multicolumn{2}{c}{Used quantities} & & & \small{References}\\
\cline{1-2}
\scriptsize{$d_z\,\,({\rm Mpc})$} & \scriptsize{3.7} &  \scriptsize{16}&  \scriptsize{76} &  \scriptsize{(1,2,3)}    \\
\scriptsize{$\theta$}\,\,({\rm degree}) & \scriptsize{30}& \scriptsize{17}&   \scriptsize{20} & \scriptsize{(4,5,6)}    \\
\scriptsize{$\epsilon_0$}\,\,({\rm TeV}) & \scriptsize{1} &  \scriptsize{1}&  \scriptsize{1} &  \scriptsize{(7,8,9)}      \\
\scriptsize{$\gamma_{\rm e,min}$} & \scriptsize{$1.2\times10^3$}&  \scriptsize{$0.67\times10^3$}&  \scriptsize{$0.5\times10^3$} & \scriptsize{(4,10,11)} \\
\scriptsize{$L_{\rm \gamma, IC}\,\, {\rm (erg/s)}$} & \scriptsize{$5\times 10^{42}$}&  \scriptsize{$5\times 10^{42}$}&  \scriptsize{$5\times 10^{42}$} & \scriptsize{(12, 13)}\\\hline
\multicolumn{2}{c}{Derived quatities} \\
\cline{1-2}
\scriptsize{$\Gamma$} & \scriptsize{7.0}&  \scriptsize{6.47}&  \scriptsize{6.29} \\
\scriptsize{$f_{\pi^0}$} & \scriptsize{$1.05\times10^{-6}$ }&  \scriptsize{ $6.68\times10^{-8}$ }&  \scriptsize{ $1.55\times10^{-8}$ } \\
\scriptsize{$\tau_{\gamma}$} & \scriptsize{$7.89\times10^{-6}$ }&  \scriptsize{ $1.40\times10^{-6}$ }&  \scriptsize{ $2.56\times10^{-7}$} \\
\scriptsize{$\gamma_{e,max}$} & \scriptsize{$3.64\times10^7$}&  \scriptsize{$3.70\times10^7$}&  \scriptsize{$3.29\times10^7$} \\
\scriptsize{$\gamma_{p,min}$} & \scriptsize{$1.0\times10^5$}&  \scriptsize{$7.0\times10^6$}&  \scriptsize{$2.4\times10^6$} \\
\scriptsize{$^a\gamma_{p,max}$} & \scriptsize{$6.73\times10^{10}$}&  \scriptsize{$6.76\times10^{10}$}&  \scriptsize{$6.04\times10^{10}$} \\
\scriptsize{$\epsilon_{\rm pk, ic}\,\, ({\rm MeV}) $ } & \scriptsize{$0.1$}&  \scriptsize{$0.5$}&  \scriptsize{$3.5$} \\
\scriptsize{$\epsilon_{\pi^0,\gamma,c}\,\,({\rm TeV})$} & \scriptsize{$0.32$}&  \scriptsize{$0.54$}&  \scriptsize{$0.05$} \\
\scriptsize{$U_B\,\, {\rm (erg/cm^3)}$} & \scriptsize{$0.52$}&  \scriptsize{$0.10$}&  \scriptsize{$0.16$} \\
\scriptsize{$U_e\,\, {\rm (erg/cm^3})$} & \scriptsize{$1.52$}&  \scriptsize{$2.41$}&  \scriptsize{$28.45$} \\
\scriptsize{$U_p\,\, {\rm (erg/cm^3})$} & \scriptsize{$2.55$}& \scriptsize{$2.69$}&  \scriptsize{$61.21$} \\
\scriptsize{$L_p\,\, {\rm (erg/s)}$} & \scriptsize{$3.74\times 10^{43}$}&  \scriptsize{$4.89\times 10^{43}$}&  \scriptsize{$2.29\times 10^{44}$} \\
\scriptsize{$L_e\,\, {\rm (erg/s})$} & \scriptsize{$1.55\times 10^{43}$}&  \scriptsize{$3.17\times 10^{43}$}&  \scriptsize{$2.27\times 10^{44}$} \\
\scriptsize{$E_{\rm p,max}\,\, {\rm (EeV)}$} & \scriptsize{$40.1$} & \scriptsize{$6.55$}&  \scriptsize{$7.92$} \\
\scriptsize{$N_{\rm \tiny UHECRs}$} & \scriptsize{$1.52$} &  \scriptsize{$0.41$}& \scriptsize{$2.63\times 10^{-6}$} \\
\hline
\hline
\end{tabular}
\end{center}
\begin{flushleft}
\scriptsize{
Notes.\\
$^a$ $\gamma_{p,max}=\frac{m_p}{m_e}\gamma_{e,max}$.\\
\textbf{References}. (1) \cite{2007ApJ...654..186F} (2) \cite{1991ApJ...373L...1T}  (3) \cite{1992ApJS...83...29S} (4) \cite{2010ApJ...719.1433A} (5) \cite{1996ApJ...467..597B}     (6) \cite{2006PASJ...58..261A}  (7) \cite{2009ApJ...695L..40A}  (8) \cite{2006Sci...314.1424A} (9) \cite{2012A&A...539L...2A}  (11)  \cite{2009ApJ...707...55A}  (10) \cite{2009ApJ...699...31A}   (12) \cite{1998A&A...330...97S}  (13) \cite{2009ApJ...690..367A}.} 
\end{flushleft}
We plot in a sky-map the 54 neutrino events  detected by the IceCube collaboration, the 72 and 27 UHECRs collected by TA and PAO experiments, respectively, and also a circular region of 5$^\circ$ around the closest RGs,  as shown in Figure \ref{Skymap}.  This figure shows that whereas there are not  neutrino track events associated to Cen A, M87 and NGC1275,  two UHECRs are only enclosed around Cen A.\\
%
%
%
By assuming that the inner part of the RG jet has the potential to accelerate particles up to UHEs, we can see that  protons  at the emitting region can achieve maximum energies of 40.1, 6.55 and 7.92 EeV for Cen A, M87 and NGC1275, respectively, as shown in Table 2.   Therefore, it is improbable that protons can be accelerated to energies as high as 57 EeV, although it is plausible for heavier accelerated ions. In this case, they can be disintegrated by infrared photons from the core. It is worth noting that any small variation in the strength of magnetic field and/or size of emitting region would allow that protons could achieve a  maximum energy  of 57 EeV for Cen A.   Similarly, supposing that the BH jet has the power also to accelerate particles  up to UHEs through Fermi processes, then during flaring intervals (for which the apparent isotropic luminosity can reach $\approx 10^{46}$ erg s$^{-1}$ and  from the equipartition magnetic field $\epsilon_B$)  the maximum particle energy of accelerated UHECRs can achieve values as high as {\small $E_{\rm max}\approx 3.0\times10^{20}\,\frac{Zq_e\,\epsilon^{1/2}_B}{\phi\, \Gamma}\,\left(\frac{L}{10^{46}\,{\rm erg/s} }\right)^{1/2}\, {\rm eV}$} \citep{2009NJPh...11f5016D}.   Describing the TeV gamma-ray spectra through p$\gamma$ interaction and extrapolating the interacting proton spectra up to energies higher than 1 EeV, we plot the UHE proton spectra expected for these three RGs (Cen A, M87 and NGC1275) with the UHECR spectra collected with PAO \citep{2011arXiv1107.4809T}, HiRes \citep{2009APh....32...53H} and TA \citep{2013ApJ...768L...1A} experiment as shown in Figure \ref{uhecr_flux}.  We can see that as energy increases the discrepancy between the UHE proton fluxes and the observed UHECR spectra increases. Taking into account the TA and PAO exposures, we estimate the number of UHECRs above 57 EeV, as shown in Table 2.  The number of UHECRs computed with our model are 1.52, 0.41 and $2.63\times 10^{-6}$ for Cen A, M87 and NGC1275, respectively. Due to extragalactic (eq. \ref{thet_EG}) and galactic (eq. \ref{thet_G}) magnetic fields, UHECRs are deflected between the true direction to the source, and the observed arrival direction. Regarding these considerations, the total deflection angle could be  as large as the mean value of $<\theta_T>\simeq$ 15$^\circ $\citep{2010ApJ...710.1422R}.  Therefore, it is reasonable to assume a circular region  with $5^\circ$  (eqs. \ref{thet_EG}) centered around each source (see fig. \ref{Skymap}). Taking into account these regions, we can see two UHECRs associated to Cen A and none to M87 and NGC1275.  We can see that number of UHECRs calculated with our model is consistent with those reported by the TA and PAO collaborations, although the  maximum proton energies are less than 57 EeV.   It is worth noting that the latter results reported by PAO suggested that UHECRs are heavy nuclei instead of protons \citep{2010PhRvL.104i1101A}. If UHECRs have a heavy composition, then a significant fraction of nuclei must survive photodisintegrations in their sources \citep{2010MNRAS.405.2810H,2009MNRAS.393.1041H}. In this case for  Z$\gtrsim$2,  the emitting region can achieve maximum energies for heavy nuclei of  $\gtrsim$ 80, 13 and 16 EeV for Cen A, M87 and NGC1275, respectively.\\ 
%
%
%
%
After fitting the TeV $\gamma$-ray spectra of the RGs with our hadronic model, from eq. (\ref{numneu1}) we obtain the neutrino fluxes and events expected in a hypothetical Km$^{3}$ neutrino telescope through the Monte Carlo simulations.  We consider a point source neutrino emitters at Cen A, M87 and NGC1275 positions with a energy range spanning from 10 GeV to 10 PeV.    The neutrino spectra are normalized from the observed TeV $\gamma$-ray spectra, assuming that the TeV $\gamma$-ray fluxes from the closest RGs are interpreted through the $\pi^0$ decay products from the p$\gamma$ interactions. The values of magnetic field, Doppler factor and emitting radius (see Table 2) were calculated as the result of fitting the SED with SSC model up to dozens of GeV.  Changes in these observables would vary the photon density generated by synchrotron radiation, and thus the number of photons scattered by inverse Compton scattering.  As neutrino fluxes were computed from the photo-hadronic interactions between Fermi-accelerated protons and the seed photons around the first and second SSC peaks, then variations in the magnetic field, Doppler factor and emitting radius would affect the target photon densities (eq. \ref{den}) and then the neutrino fluxes.  For the simulated neutrino telescope we additionally calculate the atmospheric and cosmic neutrinos expected from the portion of the sky inside a circular region centered in each source and covering $1^{\circ}$ square.  The atmospheric neutrino flux is described using the Bartol model \citep{2004PhRvD..70b3006B,2006PhRvD..74i4009B}  for the range of energy considered in this analysis. The cosmic diffuse neutrinos signal has been discussed by Waxman and Bahcall  \citep{2001PhRvD..64b3002B, wax98} and the upper bound for this flux is $E^{2}_{\nu}\,d\Phi/dE_{\nu} <2\times 10^{-8}\, {\rm GeV\,cm^{-2}\,s^{-1}\,sr^{-1}}$ \citep{1999PhRvD..59b3002W}.  We rule out the atmospheric muon ``background'' from this analysis due to earth filtration and to the softener spectrum with respect to the neutrino signal and ``backgrounds''  considered.  We plot the neutrino spectra expected for CenA, M87 and NGC1275, as shown in figure \ref{neutr_flux}.  In this figure we have considered Fermi-accelerated protons interacting with the seed photons around the first and second SSC peaks, for this reason neutrino spectra are exhibited, firstly, by declining power laws and after by peaking around 1 EeV as  expected \citep{2008PhRvD..78b3007C}.    By computing the signal to noise ratio for one year of observation in  Km$^3$ neutrino telescope, the neutrino IceCube flux \citep{2014PhRvL.113j1101A} and the upper limits set by IceCube (IC40; \cite{2011PhRvD..83i2003A}), PAO \citep{2011PhRvD..84l2005A}, RICE \citep{2012PhRvD..85f2004K} and ANITA \citep{2010PhRvD..82b2004G} we confirm the impossibility to observe HE and UHE neutrinos from RGs. No visible neutrino excess (under the assumption of p$\gamma$ interaction model) respect to atmospheric and cosmic neutrinos is expected  for the three RGs considered, as it is shown in figure \ref{neut_event}.  In our model,  the proton and neutrino spectra are normalized with the TeV $\gamma$-ray fluxes from the closest RGs, therefore to obtain the diffuse flux from other RGs would be needed to model the TeV $\gamma$-ray flux of each RG  which is outside of the scope of this paper.   It is important to say that if UHECRs are heavy as suggested by PAO \citep{2010PhRvL.104i1101A},  HE neutrinos from UHE nuclei are significant lower than the neutrino flux obtained by UHE protons \citep{2010PhRvD..81l3001M}.
%
\vspace{1cm}
\section{Summary and conclusions}
We have proposed a leptonic and hadronic model to explain the broadband  SED spectrum observed in the closest RGs. In the leptonic model, we have used the SSC emission to describe the SED up to dozens of GeV. To explain the spectrum from hundreds of GeV up to a few TeV, we have introduced the hadronic model assuming that  accelerated protons in the inner jet interact with  the photon population at the SED peaks. Evoking these interaction,  we have interpreted the TeV $\gamma$-ray spectra as $\pi^0$ decay products in Cen A, M87 and NGC 1275.\\
Correlating the TeV $\gamma$-ray and HE neutrino fluxes through p$\gamma$ interactions, we have computed the HE and UHE neutrino fluxes, and the neutrino event rate expected in a kilometric scale neutrino detector.   The neutrino event rate was obtained through MC simulation by considering a region of 1$^{\circ}$ around the source position and assuming a hypothetical Km$^3$ Cherenkov telescope.  We found that the neutrino fluxes produced by p$\gamma$ interactions close to the core of RGs cannot explain the astrophysical flux and the expected $\nu_{\mu}$ events in a neutrino telescope are consistent with the nonneutrino track-like associated with the location of the closest RGs \citep{2013PhRvL.111b1103A, 2013Sci...342E...1I, 2014PhRvL.113j1101A, 2015ATel.7856....1S}. In addition, the atmospheric muon neutrino background is also shown.\\
Extrapolating the proton spectrum by a simple power law up to UHEs, we have computed the number of UHECRs expected from Cen A, M87 and NGC1275.  We found that those UHECRs obtained with our model is in agreement with the TA and PAO observations.   Although UHECRs from Cen A can hardly be accelerated up to the PAO energy range at the emitting region (E$_{max}$= 40  EeV), they could be accelerated during the flaring intervals, with small changes in the strength of magnetic field and/or emitting region and in the giant lobes.  It is very interesting the idea that UHECRs could be accelerated partially in the jet at energies ($<40\times 10^{19}$ eV) and partially in the Lobes at ($E>40\times 10^{19}$ eV) \citep{2014ApJ...783...44F}.  If UHECRs have a heavy composition as suggested by PAO \citep{2010PhRvL.104i1101A}, UHE heavy nuclei in Cen A   could be accelerated  at energies greater than $\sim$ 80 EeV, thus reproducing the detections  reported by PAO.    It is worth noting that if radio Galaxies are the sources of UHECRs, their on-axis counterparts (i.e. blazars, and flat-spectrum radio quasars) should be considered a more powerful neutrino emitters \citep{2001PhRvL..87v1102A, 2014JHEAp...3...29D,2014PhRvD..90b3007M}.  In fact, a PeV neutrino shower-like was  recently associated with the flaring activity of the blazar  PKS B1424-418 \citep{2016arXiv160202012K}. \\
In summary,  we have showed that leptonic SSC and hadronic processes are required to explain the $\gamma$-ray fluxes at GeV- TeV energy range. We have successfully described the TeV $\gamma$-ray  \citep{2009ApJ...695L..40A, 2006Sci...314.1424A, 2010ApJ...710..634A}, HE neutrinos \citep{2013PhRvL.111b1103A, 2013Sci...342E...1I, 2014PhRvL.113j1101A, 2015ATel.7856....1S} and UHECRs \citep{2011arXiv1107.4809T, 2013ApJ...768L...1A} around the closest RGs.
\acknowledgments
We thank the anonymous referee  for a critical reading of the paper and valuable suggestions that helped improve the quality and clarity of this work.  We also thank to Charles Dermer,  Tyce DeYoung,  Markus B\"{o}tcher,  Bin Zhang  for useful discussions.  
%
%
%

%
%
\newpage
%
%
%
\appendix 
\section{Tools for fit}
The values reported in Table 1 (break energies, spectral index and normalizations) for synchrotron, Compton scattering and p$\gamma$ spectra were obtained as follows.
\paragraph{Synchrotron spectrum}. We require the synchrotron spectrum (eq. \ref{espsyn}) with $ A_{syn,\gamma}=[0]$, $\alpha_e=[1]$, $ \epsilon^{syn}_{\gamma,m} =[2]$ and $\epsilon^{syn}_{\gamma,c} =[3]$. Then, it   is written as 
{\small
\begin{equation}
\label{espsyn_0}
\left[\epsilon^2_\gamma N(\epsilon_\gamma)\right]_{\gamma, syn} = [0]
\cases {
\left(\frac{\epsilon_\gamma}{[2]}\right)^\frac43    &  $\epsilon_\gamma < [2]$,\cr
 \left(\frac{\epsilon_\gamma}{[2]}\right)^{-\frac{[1]-3}{2}}  &  $[2] < \epsilon_\gamma < [3]$,\cr
\left(\frac{[3]}{[2]}\right)^{-\frac{[1]-3}{2}}    \left(\frac{\epsilon_\gamma}{[3]}\right)^{-\frac{[1]-2}{2}},           &  $[3] < \epsilon_\gamma$\,. \cr
}
\end{equation}
\small}
\paragraph{Compton scattering  spectrum}. Given the Compton scattering  spectrum (eq. \ref{espic}) and doing $A_{ssc,\gamma}=[4]$, $ \epsilon^{ssc}_{\gamma,m}=[5]$ and $\epsilon^{ssc}_{\gamma,c}=[6]$, this spectrum is in the form
{\small
\begin{equation}
\label{espic_0}
\left[\epsilon^2_\gamma N(\epsilon_\gamma)\right]_{\gamma, ssc} = [4]
\cases {
\left(\frac{\epsilon_\gamma}{[5]}\right)^\frac{4}{3}    &  $\epsilon_\gamma < [5]$,\cr
 \left(\frac{\epsilon_\gamma}{[5]}\right)^{-\frac{[1]-3}{2}}  &  $[5] < \epsilon_\gamma < [6]$,\cr
\left(\frac{[6]}{[5]}\right)^{-\frac{[1]-3}{2}}    \left(\frac{\epsilon_\gamma}{[6]}\right)^{-\frac{[1]-2}{2}},           &  $[6] < \epsilon_\gamma$\,.\cr
}
\end{equation}
\small}
\paragraph{P$\gamma$ spectrum}. Considering  p$\gamma$ spectrum  (eq. \ref{pgammam}) with $A_{p\gamma} =[7]$  and $\alpha_p=[8]$, then it can be written as   
%
{\small
\bary
\label{pgammam_0}
\left[\epsilon^2_\gamma N(\epsilon_\gamma)\right]_{\gamma, \pi^0} =  [7] \cases{
\left(\frac{\epsilon^{\pi^0}_{\gamma,c}}{\epsilon_{0}}\right)^{-1} \left(\frac{\epsilon_{\gamma}}{\epsilon_{0}}\right)^{-[8]+3}          &  $ \epsilon_{\gamma} < \epsilon^{\pi^0}_{\gamma,c}$\cr
\left(\frac{\epsilon_{\gamma}}{\epsilon_{0}}\right)^{-[8]+2}                                                                                        &   $\epsilon^{\pi^0}_{\gamma,c} < \epsilon_{\gamma}$\,.\cr
}
\eary
}
We use  the method of Chi-square $ \chi^2$ minimization as implemented in the ROOT software package \citep{1997NIMPA.389...81B}. 
\end{document}